\documentclass[conference, twocolumn, 10pt]{IEEEtran}

\ifCLASSINFOpdf
  \usepackage[pdftex]{graphicx}
   \graphicspath{{..1/pdf/}{../jpeg/}}
   \DeclareGraphicsExtensions{.pdf,.jpeg,.png}
\else
   \usepackage[dvips]{graphicx}
   \graphicspath{{../eps/}}
   \DeclareGraphicsExtensions{.eps}
\fi

\usepackage[noadjust]{cite}
\usepackage{comment} 
\usepackage[cmex10]{amsmath}
\usepackage{amsfonts, amsmath, amsthm, amstext, amssymb, mathrsfs, graphicx, color, url}
\usepackage{subcaption}
\usepackage{array}
\usepackage{mdwmath}
\usepackage{mdwtab}
\usepackage{algpseudocode}
\usepackage{algorithm}
\usepackage{url}
\usepackage{setspace}
\usepackage{footnote}
\usepackage[skip=0pt]{caption}
\usepackage{eqparbox}
\usepackage{fixltx2e}

\def\IC{{\mathbb C}}

\def\IR{{\mathbb R}}

\def\calA{{\mathcal A}}
\def\calB{{\mathcal B}}
\def\calC{{\mathcal C}}
\def\calD{{\mathcal D}}

\def\calR{{\mathcal R}}
\def\calS{{\mathcal S}}

\def\calU{{\mathcal U}}

\def\calX{{\mathcal X}}

\def\bA{{\pmb A}}

\def\bH{{\pmb H}}
\def\bI{{\pmb I}}

\def\bW{{\pmb W}}

\def\ba{{\pmb a}}

\def\bf{{\pmb f}}

\def\bh{{\pmb h}}

\def\br{{\pmb r}}

\def\bv{{\pmb v}}
\def\bw{{\pmb w}}
\def\bx{{\pmb x}}

\def\bz{{\pmb z}}

\newcommand{\bPsi}{  \pmb{\Psi}  }

\newcommand{\upperRomannumeral}[1]{\uppercase\expandafter{\romannumeral#1}}

\newcommand{\st}{ \textup{s. t.} }
\newcommand{\argmin}{ \textup{argmin} }

\newcommand{\lrb}[1]{ \lbrace #1 \rbrace }

\newtheorem{lemma}{Lemma}

\newtheorem{proposition}{Proposition}

\newtheorem{remark}{Remark}

\newcommand{\bom}{  \pmb{\omega}  }

\begin{document}

\title{Coordination and Antenna Domain Formation in Cloud-RAN systems}

\author{\IEEEauthorblockN{Hadi Ghauch\IEEEauthorrefmark{1},
Muhammad Mahboob Ur Rahman\IEEEauthorrefmark{1}, 
Sahar Imtiaz\IEEEauthorrefmark{1},
James Gross\IEEEauthorrefmark{1},  
}

\IEEEauthorblockA{\IEEEauthorrefmark{1}School of Electrical Engineering and the ACCESS Linnaeus Center, Royal Institute of Technology (KTH) }

}

\maketitle

\begin{abstract}
We study here the problem of Antenna Domain Formation (ADF) in cloud RAN systems, whereby multiple  remote radio-heads (RRHs) are each to be assigned to a set of antenna domains (ADs), such that the total interference between the ADs is minimized.  
We formulate the corresponding optimization problem, by introducing the concept of \emph{interference coupling coefficients} among pairs of radio-heads. We then propose a low-overhead algorithm that allows the problem to be solved in a distributed fashion, among the aggregation nodes (ANs), and establish basic convergence results. Moreover, we also propose a simple relaxation to the problem, thus enabling us to characterize its maximum performance. We follow a layered coordination structure: after the ADs are formed, radio-heads are clustered to perform coordinated beamforming using the well known Weighted-MMSE algorithm. Finally, our simulations show that using the proposed ADF mechanism would significantly increase the sum-rate of the system (with respect to random assignment of radio-heads). 
\end{abstract}

\begin{IEEEkeywords}
5G, Cloud RAN, radio head assignment, antenna domain formation, interference coupling, block coordination descent 
\end{IEEEkeywords}

\IEEEpeerreviewmaketitle

\section{Introduction}

The \emph{Cloud-Radio Access Network (C-RAN)} is identified as one of the promising architectures to address the challenges of 5G systems, namely, the requirement for high spectral efficiency within a particularly dense deployment (both users and access nodes)~\cite{METISD62}. 
C-RAN systems are characterized as a centralized solution for interference coordination: remote radio-heads (RRH) act as access nodes, and their baseband processing capabilities vary from full digital signal processing capability (i.e., base stations), to `dumb antennas' with no baseband capabilities (such as distributed MIMO systems). Such radio-heads are connected via high-capacity (possibly wireless) links to so-called \emph{aggregation nodes (ANs)}, each essentially acting as a large processing unit. Thus, such architectures are natural candidates for interference coordination.

In dense deployments, coordination among base stations was identified as the key to achieving high spectral efficiency: indeed the ideas of Coordinated Multi-point (CoMP)~\cite{gesbert_comp_10},~\cite{Emil:TSP:2011}  and Interference Alignment (IA)~\cite{cadambe_interference_2009},~\cite{maddah-ali_communication_2008} were central to achieve higher spectral efficiency. However, when applied to conventional cellular systems, such techniques have the stringent requirement that they need to be distributed, i.e., to only use local CSI at each node: the overhead associated with such techniques has been identified as a (potentially) limiting factor of the sum-rate gains brought about by techniques such as IA~\cite{ayach_overhead_12},~\cite{ayach_overhead_11},~\cite{peters_user_2012} and \cite{Lozano_limits_coop_13}. This essentially puts hard limits on the effectiveness of the latter techniques. However, this limitation is lifted in the C-RAN architecture, since ANs can be assumed to have perfect CSI of a large area, and can perform coordination in a centralized manner. Earlier related work has been reported in~\cite{Yu_limited_BH} where the authors investigate the beamforming design problem (for sum-rate maximization), in the context of cellular systems with limited backhaul capacity.

In ~\cite{Lee_BS_clustering_CRAN_13}, the authors study the problem of dynamic clustering in dense deployments (for joint transmission), by characterizing the statistics of the instantaneous signal-to-interference-and-noise ratio (SINR), via tools from stochastic geometry.  In our earlier paper~\cite{Rahman_RRHclust_15}, we investigated radio-head coordination (namely coordinated beamforming), in a typical Cloud RAN setup, with a large number of radio-heads and users, served by one AN. In this work, however, we look higher into the coordination hierarchy, by investigating the so-called \emph{Antenna Domain Formation (ADF)} problem: given a set of radio-heads (each serving a set of users), and a set of ANs, what is the best assignment of radio-heads to ANs, such that the total interference leakage between the ADs, is minimized. 
Studying the latter setup is the main contribution and novelty of this paper. In that sense, we formulate the ADF problem as integer programming problem, and devise an iterative algorithm for solving it. We also relax the latter problem to obtain a lower bound on the maximum performance of our algorithm. Moreover, we investigate the effect of using a layered coordination structure, whereby further coordination mechanism (coordinated beamforming) are put in place. We underline the that fact that this work is currently being extended to a journal form\cite{Ghauch_ADF_journ}.

In the following, we use bold upper-case letters to denote matrices, bold lower-case to denote vectors, and calligraphic letters to denote sets. Furthermore, for a given matrix $\pmb{A}$, $[\pmb{A}]_{i:j}$ denotes the matrix formed by taking columns $i$ to $j$, of $\pmb{A}$,  $\Vert \pmb{A} \Vert_F^2$ its Frobenius norm, $|\pmb{A}|$ its determinant, $\bA^T$ its transpose, and $\pmb{A}^\dagger$ its conjugate transpose . $ [\bA]_{i,j} = a_{i,j}$ denotes element $(i,j)$ in a matrix $\bA$, and $[\ba]_i$ element $i$ in a vector $\ba$. While $\bI_n$ denotes the $n \times n$ identity matrix, 
  $\pmb{1}_n$ denotes the $n \times 1$ vector of ones, 
  $\calB_N$ denotes space of $N$-dimensional binary vectors,
  and $ \Pi_{\calS}[\bx]$ is the Euclidean projection of a vector $\bx$, into some (possibly non-convex) set $\calS$.

\section{System Model}
\begin{figure}[t]
  \center
  \includegraphics[scale=.6]{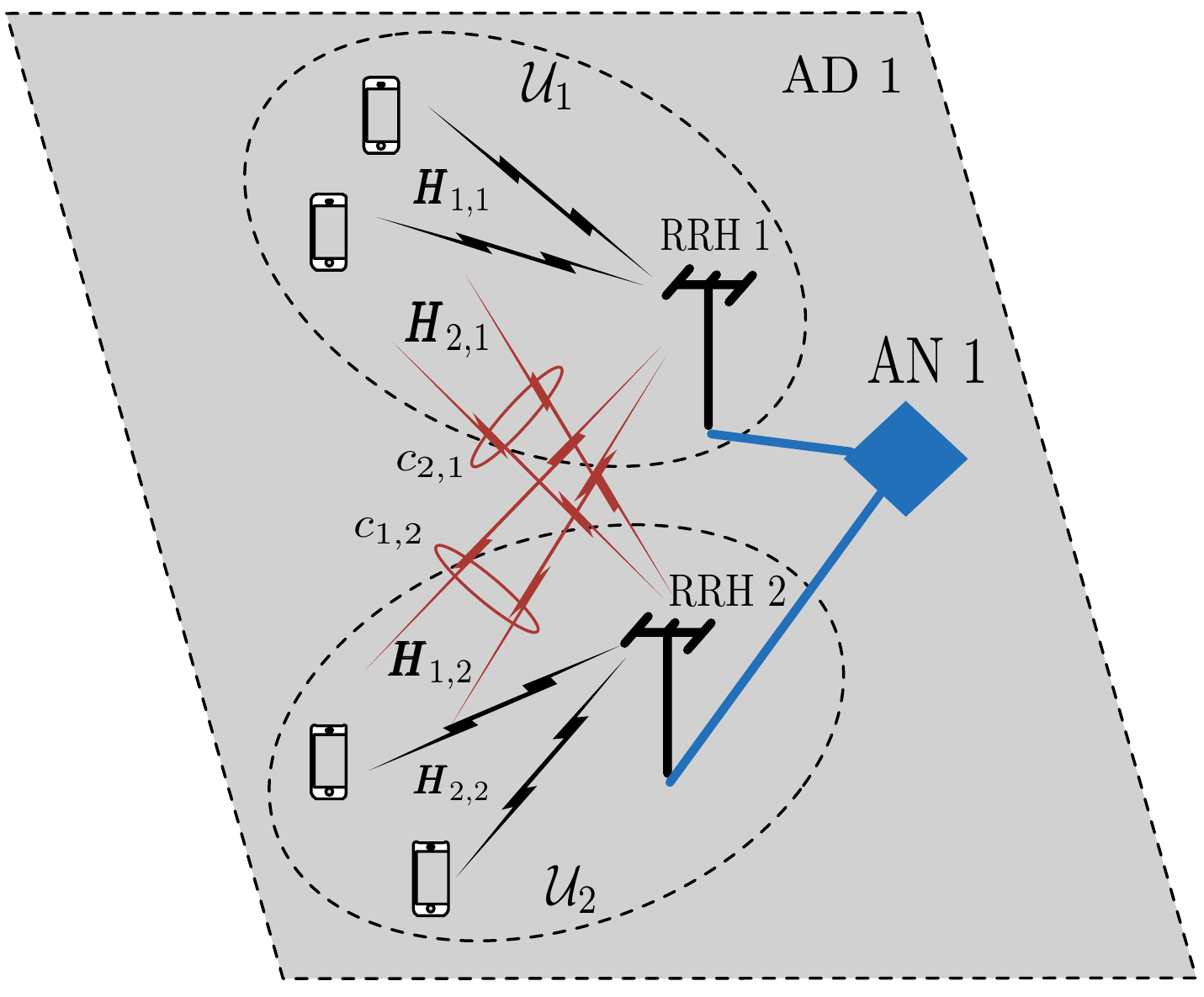}
  \caption{Toy Example }
    \label{fig:toyex}
\end{figure}
Given a large area of interest, i.e. a spatial area of a certain size, with $N$ of RRHs and $A$ ANs, we wish to address the issue of ADF, i.e. which radio heads should be assigned to which AN. In other words, given a set $\calR$ of radio heads in an area, the problem is to assign them to a set of ADs (where each AD is controlled by one AN) $\calA$ , such that total interference coupling between the ADs is minimal. Let $\calA$ denote a set of ANs (where $A \triangleq |\calA| $), and $\calR$ the set of radio-heads where $N= |\calR| $. 

We assume that radio head $ i \in \calR $ is equipped with $M_i$ antennas, and serving users which are single-antenna receivers (we assume single antenna users for simplicity of exposition, though this can be extended to multi-antenna receivers).   
Let $\calU_i$ be the set of users served by radio head $i \in \calR $. Then, $\bH_{i,j} \in \IC^{ |\calU_j| \times M_i }  $  denotes the channel from the antennas of radio head $i \in \calR $, to the users served by radio head $j \in \calR$, i.e. $\calU_j$, and $\bW_{i} \in \IC^{M_i \times |\calU_i|  } $ be the precoder that RRH $i$ employs to serve its users $\calU_i$. A small toy example is illustrated in Fig.~\ref{fig:toyex}.

This work essentially addresses the interference between ADs, as this type of interference limits significantly the performance of the entire system. It is clear that for any form of operation within the ADs, as well as for any type of interference mitigation technique applied between the ADs, there are more suitable and less suitable AD choices for the radio heads.

We assume that the load has been pre-allocated among all ANs, implying that the $k$th AN will have a load of $\gamma_k \geq 0$. A special case of this is when the load is equally distributed across all ANs, i.e., $\gamma_k$ is the same among all the ANs. Furthermore, each RRH is to be assigned to one AN only. 
Moreover, it is assumed that each AN has perfect CSI of the users to which it is associated. 

Hence, the main idea in order to form ADs is to reduce the interference coupling between the different ADs. Denote by $\alpha_{i,j} \in \mathbb{R}_+$ the interference leakage between radio head $i \in \calR$ and radio head $j \in \calR$, and note that this denotes the leakage in the direction of $j$, i.e. $\alpha_{i,j}$ is a measure that represents the  interference caused by RRH $i$ on terminals associated to RRH $j$, if $i$ and $j$ are assigned to be in different ADs. Due to asymmetric channel gains and terminal associations, notice in particular that $\alpha_{i,j} \neq \alpha_{j,i}$.

Thus, we define a coupling coefficient $c_{i,j}$ as  the interference coupling coefficient between RRH $i$ and RRH $j$. The latter  can be then viewed as the cost of having both radio heads in the same AD. Thus, one intuitive choice for the latter is to select $c_{i,j}$ as the total interference leakage between the two radio heads, i.e., 
\begin{align}
 c_{i,j} = \alpha_{i,j}+\alpha_{j,i} = \Vert \bH_{i,j}\bW_j \Vert_F^2 + \Vert \bH_{j,i}\bW_i \Vert_F^2 
\end{align}
However, incorporating the precoders into the coupling coefficients makes the overall system quite challenging since the coupling coefficients have to be updated quite frequently. Thus, we formally define the latter quantity as,
\begin{align} \label{eq:cij_inst}
c_{i,j} = 
\begin{cases} 
    \Vert \bH_{i,j} \Vert_F^2 + \Vert \bH_{j,i} \Vert_F^2 , \ \forall \ i \neq j \\
    \  0 \ , \forall i = j      
\end{cases}
\end{align}
Let $\bPsi$ be the matrix formed by gathering all those coefficients, i.e. $[\bPsi]_{i,j} = c_{i,j} , \ \forall \ (i,j) \in \calR^2$, where $\bPsi \in \IR_+^{N \times N} $ is such that $\bPsi = \bPsi^T $.

Furthermore, denote by $x_{i,k} \in \{0,1\}$ the binary decision variable if RRH $i$ is assigned to AD $k$ (or not), and denote by $\bx_k \in \lrb{0,1}^N $ the assignment vector of AN $k \in \calA$. We formulate the ADF problem as the following integer programming problem:
\begin{align*} 
(P1)\begin{cases} 
   \min f(\lrb{x_{i,j} } ) = \sum_{ k =1 }^A \sum_{l \neq k}^A  \left( \sum_{i=1}^N \sum_{j\neq i}^N  c_{i,j} x_{i,k} x_{j,l} \right)  \\
   \st \ \sum_{i=1}^N \beta_{i,k} x_{i,k} = \gamma_k , \ \forall k \in \calA \\
   \hspace{.7cm} \sum_{k=1}^A x_{i,k} \leq 1 , \ \forall i \in  \calR \\
   \hspace{.7cm}  x_{i,k} \in \lrb{0,1} , \ \forall (i,k) \in \calA \times \calR
\end{cases}
\end{align*}
Note that one can rewrite the above problem in terms of $\lrb{\bx_k}$ as follows,
\begin{align*} 
(P2)\begin{cases} 
   \min f(\lrb{ \bx_k } ) = \sum_{ k =1 }^A \sum_{l \neq k}^A  \bx_k^T \bPsi \bx_l   \\
   \st \  \ \sum_{k=1}^A \bx_{k} \leq  \pmb{1}_N  \\
   \hspace{.7cm}   \ \pmb{\beta}_k^T  \bx_k = \gamma_k , \  \bx_k \in  \calB_N , \  \ \forall k \in \calA \\   
\end{cases}
\end{align*}
where the above inequality holds element-wise. In the above, $\sum_{k=1}^A \bx_{k} \leq  \pmb{1}_N$ is the \emph{assignment constraint} forcing that each RRH is assigned to at most one AD. Moreover,  denoting by $ \pmb{\beta}_k \in \mathbb{R}_+^N $ the set of loading factors for AD $k$ (such that $ \pmb{1}^T \pmb{\beta}_k =1  , \ \ \forall k \in \calA $),  $\pmb{\beta}_k^T  \bx_k = \gamma_k$ is the \emph{loading constraint} for the $k$th AD (i.e. the load profile for AD $k$).  Using this equivalent form, it becomes evident that $f$ indeed is not jointly convex in all the variables, due to the coupling among $\bx_k$ and $\bx_l$. However, this does reveal an inherent  bi-linear structure of $f$ (taken separately in each variable, $f$ is linear)  that we exploit for the optimization. 
Intuitively, $f$ models the sum-total coupling that exists \emph{between} the individual ADs. 

\section{Proposed Algorithm }
\subsection{Block Coordinate Descent}
The coupling among the variables in $f(\bx_1, ..., \bx_A)$,  makes $(P2)$ a perfect candidate for a Block-Coordinate Descent (BCD) approach (also known as the Gauss-Siedel method). In a nutshell, BCD works iteratively, by fixing all variables but one block, i.e. fix $(\bx_1, ..., \bx_{k-1} , \bx_{k+1}, ..., \bx_A )$, and optimize for $\bx_k$, iteratively. 
Let $n$ denote the iteration number, i.e., $\bx_k^{(n)}$ denotes the value of $\bx_k$ at the $n$th iteration. At the $n$th iteration, given that $ (\bx_{1}^{(n)},..., \bx_{k-1}^{(n)}, \bx_{k+1}^{(n)} ,..., \bx_{A}^{(n)} )$ are fixed,  $\bx_k$ is optimized accordingly.

\begin{align} \label{eq:bcdit}
\lrb{ \bx_k^{(n+1)} } &= \underset{\bx_k}{\argmin} \ f(\bx_{1}^{(n+1)},..., \bx_{k-1}^{(n+1)}, \bx_k , \bx_{k+1}^{(n)} ,..., \bx_{A}^{(n)}) , \nonumber \\
 &=  \underset{\bx_k}{\argmin} \ f( \bx_k, \bz_k^{(n)} ) , \ \forall k \in \calA  
 \end{align}
 where $$\bz_k^{(n)} = (\bx_{1}^{(n+1)},..., \bx_{k-1}^{(n+1)}, \bx_{k+1}^{(n)} ,..., \bx_{A}^{(n)} )$$ denotes the block of fixed variables in the above BCD iteration. Stated differently, in the above,  $\bx_{k-1}^{(n+1)}$ indicates that block $\bx_{k-1}$ (for instance) has been already updated, while block $\bx_{k+1}$ hasn't. 
Thus, as seen from~\eqref{eq:bcdit}, BCD generates a sequence of iterations $ \lrb{\bx_k^{(n)}} $ that converge to a limit point (this will be formalized later in this section). 
Moreover, we note that $f( \bx_k, \bz_k^{(n)} )$ (which denotes the function $f(\bx_k)$, when the variables in block $\bz_k^{(n)}$ are fixed), can be rewritten as, 
\begin{align} \label{eq:rk}
&f( \bx_k, \bz_k^{(n)} ) = \bx_k^T \bPsi  \left(  \sum_{l = 1}^{k-1}  \bx_l^{(n+1)} + \sum_{l = k+1}^A \bx_l^{(n)} \right) \triangleq \bx_k^T \br_k^{(n)}  \nonumber \\
& \textrm{where} \ \br_k^{(n)} \triangleq \bPsi  \left(  \sum_{l = 1}^{k-1}  \bx_l^{(n+1)} + \sum_{l = k+1}^A \bx_l^{(n)} \right) 
\end{align}
is referred to as the residual of AD $k$, at the $n$th iteration. Moreover, we clearly see that $f( \bx_k, \bz_k^{(n)} )$  is linear, implying that when all but one block are fixed, $f$ is linear.
This indeed shows that $(P2)$ is equivalent to a series of problems that are solved iteratively, and block-by-block. 
Now that we have described the BCD framework, we focus on the solution of the optimization problem in~\eqref{eq:bcdit},   within each BCD iteration.

With that in mind, the update for $\bx_k$, i.e. $\bx_k^{(n+1)}$ in~\eqref{eq:bcdit}, is,
\begin{align} \label{opt:xk_opt} 
\bx_k^{(n+1)} = \begin{cases} 
   \underset{ \bx_k }{\argmin}  \ f(\bx_k,  \bz_k^{(n)} )    \\
   \st \ \pmb{\beta}_k^T \bx_k = \gamma_k , \ \ \bx_{k} \leq \bom_k^{(n)} ,  \  \ \bx_k \in  \calB_N  
\end{cases}
\end{align}
where $\bom_k^{(n)} $ is the vector of \emph{residual assignments} dependent on the the assignments of all other ADs. 
\begin{align} \label{eq:wk}
  \bom_k^{(n)} \triangleq \pmb{1}_N - \left(  \sum_{l = 1}^{k-1}  \bx_l^{(n+1)} + \sum_{l = k+1}^A \bx_l^{(n)} \right)
\end{align}
The above problem is known as a mixed integer linear program, and is NP-hard due to its inherent combinatorial nature. However, there exists many polynomial-time approximation algorithms that, for all practical purposes, solve it globally (such efficient solvers are found in MATLAB and CVX). 

It becomes clear at this stage that BCD transforms $(P2)$ into a series of $A$ parallel subproblems, where each   can be solved in a distributed way, i.e. locally at each AN.

\subsection{Algorithm Description }
The use of BCD for solving $(P2)$ goes hand in hand with making the problem naturally decoupled: when $ \lrb{\bx_l}_{l\neq k}$ are fixed, the cost function decouples in $\bx_k$ can thus be solved separately by AD $k$, without any loss in optimality. In that sense, the optimal update for $\bx_k$ at AD $k$, depends on the assignments at all the other ADs, that have to be shared. 
Given assignments from other ADs, $(\bx_{1}^{(n+1)},..., \bx_{k-1}^{(n+1)}, \bx_{k+1}^{(n)} ,..., \bx_{A}^{(n)} )$, AD $k$ forms the residual  $\br_k^{(n)}$, and can proceed to solve its optimization problem locally, and update $\bx_k^{(n+1)}$. The process is formalized in Algorithm~\ref{alg:1}. 


\begin{algorithm} 
\caption{ADF via BCD} \label{alg:1}
\begin{algorithmic}
\State \textbf{Input:} $ \bPsi, \ N , \ A $  
\For{$n=0,1, \cdots , L-1$}
\State // \emph{procedure at each AN }
\State obtain   $(\bx_{1}^{(n+1)},..., \bx_{k-1}^{(n+1)}, \bx_{k+1}^{(n)} ,..., \bx_{A}^{(n)} )$  at AD $k$
\State compute residual $\br_k^{(n)}$, using~\eqref{eq:rk}
\State compute residual assignment $\bom_k$, using~\eqref{eq:wk} 
\State compute $\bx_k^{(n+1)}$ as solution to~\eqref{opt:xk_opt}
\EndFor 
\State \textbf{Output:} $ \lrb{ \bx_1^{(L)} , ..., \bx_A^{(L)} } $  
\end{algorithmic}
\end{algorithm}

\subsection{Convergence} \label{sec:conv}
Let $\lrb{ \bx_k^{(n)} } $  be the sequence iterates produced by the BCD in~\eqref{eq:bcdit}, and $ \lrb{ \bx_k^\star }  \triangleq \lim_{n \rightarrow \infty}  \lrb{ \bx_k^{(n)} }  $. The monotonic nature of the BCD iterations is established below.  

\begin{lemma}[Monotonicity] \label{lem:mono}
 With each update $ \bx_k^{(n)}  \rightarrow \bx_{k+1}^{(n)} $, $f$ is non-increasing. Moreover, the sequence of function iterates $ \lrb{ f(\bx_1^{(n)}, ..., \bx_A^{(n)} ) }_n $ converges to a limit point $f( \lrb{ \bx_k^\star }) $.    
\end{lemma}
\begin{IEEEproof} Refer to Appendix~\ref{sec:app1}
\end{IEEEproof}
Although the above result establishes the convergence of the proposed BCD method, it only establishes convergence to a limit.

\begin{remark} \rm \label{rem:conv}
Although the updates generated by the BCD iteration are shown to converge monotonically to a limit point, it cannot be established that the latter corresponds to a stationary point of $f$, namely due to 
\begin{itemize}
 \item[-] the presence of the binary constraint, that prevents the use of advanced BCD convergence results such as~\cite{Tseng_convBCD_01} 
 \item[-] the coupling in the assignment constraint, i.e.  $\bx_1 + ... + \bx_A \leq \pmb{1}_N $, cannot be handled by standard BCD convergence results. 
\end{itemize}

\end{remark}

\subsection{Performance bounds} \label{sec:lb}
Here we attempt to shed light on the maximum performance that Algorithm 1 can deliver. This is achieved by relaxing the original problem in $(P2)$. One of the most well known relaxations for problems such as $(P2)$ is done by relaxing the binary constraint on $\bx_k$. In this case however, we note that the loading constraint, $ \pmb{\beta}_k^T  \bx_k = \gamma_k  $, does not make much sense: a quick look at this case reveals that the loading constraint makes some problems infeasible (this is expected  since it is only effective when $\bx$ is binary). We thus conclude that a sensible relaxation  has to involve both the binary constraint, and the loading constraint. The resulting problem becomes,
\begin{align*} 
(P3)\begin{cases} 
   \min f(\lrb{ \bw_k } ) = \sum_{ k =1 }^A \sum_{l \neq k}^A  \bw_k^T \bPsi \bw_l   \\
   \st  \  \sum_{k=1}^A \bw_{k} \leq \pmb{1}_N ,  \\
        \hspace{.6cm} \bw_k \in [0, \ 1]^N , \  \ \forall k \in \calA \\
\end{cases}
\end{align*}
where $ \lrb{ \bw_k}$, the optimization variables are no longer binary. 
The fact that the optimal solution of $(P3)$ is a lower bound on the original problem $(P2)$, follows immediately from the relaxation arguments. Despite its simple form, globally solving the above problem is not straightforward, namely due to coupling among the variables, $\lrb{\bw_k}$, and that $f$ is not convex in $\lrb{\bw_k}$ (since $\bPsi$ is not positive-definite). However, we recall that the same BCD procedure that was used to solve $(P2)$, can be applied to the relaxed problem. Thus, the sequence of iterates generated by the BCD is given by, 
\begin{align} \label{opt:wk_opt} 
\bw_k^{(n+1)} = \begin{cases} 
   \underset{ \bw_k }{\argmin}  \ f(\bw_k,  \bz_k^{(n)} )    \\
   \st \ \bw_{k} \leq \bom_k^{(n)} ,  \ \ \bw_k \in [0, \ 1]^N
\end{cases}
\end{align}
where $f( \bw_k, \bz_k^{(n)} ) $ and $\bom_k^{(n)} $ are given as follows, 
\begin{align} \label{eq:fk}
f( \bw_k, \bz_k^{(n)} ) &= \bw_k^T \bPsi  \left(  \sum_{l = 1}^{k-1}  \bw_l^{(n+1)} + \sum_{l = k+1}^A \bw_l^{(n)} \right) \triangleq \bw_k^T \br_k^{(n)} \nonumber \\
\bom_k^{(n)} &= \pmb{1}_N - \left(  \sum_{l = 1}^{k-1}  \bw_l^{(n+1)} + \sum_{l = k+1}^A \bw_l^{(n)} \right) 
\end{align}

Let $\lrb{ \bw_k^{(n)} } $  be the sequence iterates produced by the BCD in~\eqref{opt:wk_opt}, and $ \lrb{ \bw_k^\star }  \triangleq \lim_{n \rightarrow \infty}  \lrb{ \bw_k^{(n)} }  $. Then, by applying BCD to relaxed problem $(P3)$, yields the desired lower bound on $(P2)$. Moreover, the resulting solution to $(P3)$, $ \lrb{\bw_k^\star }$ is such that $ f( \lrb{ \bx_k^\star }) \geq  f( \lrb{ \bw_k^\star })  $. We recall at this stage that $\lrb{\bw_k^\star }$ doesn't necessarily correspond to an assignment variable. Thus, 
a natural question is whether one can obtain an estimate of the solution to $(P2)$, $ \lrb{\hat{\bx}_k^\star}$, from the solution of the relaxed problem $ \lrb{\bw_k^\star } $. The relation between $\hat{\bx}_k^\star$ and ${\bw}_k^\star$  can be formalized as follows, 
\begin{align}
&\hat{\bx}_k^\star = \Pi_{\calD_k}[ \bw_k^\star ] , \ \forall k \in \calA  \nonumber  \\
&\textrm{where } \calD_k = \lrb{ \bw \  | \  \bw \in \calB_N , \ \pmb{\beta}_k^T \bw = \gamma_k    }
\end{align}
Stated differently, $\hat{\bx}_k^\star$ is the Euclidean projection of $\bw_k^\star$, on the non-convex set $\calD_k$. However, a closer look at the latter reveals that the projection in this case, does 
not always yield a unique point $ \hat{\bx}_k^\star$. Consequently, although the solution of the relaxed problem, $\bw_k^\star$, cannot be used as a basis for assignment, we indeed use it as a lower bound on the cost function value that Algorithm 1 yields.

\subsection{System-level Operation}

\begin{figure*}%
\centering
\includegraphics[scale=.6]{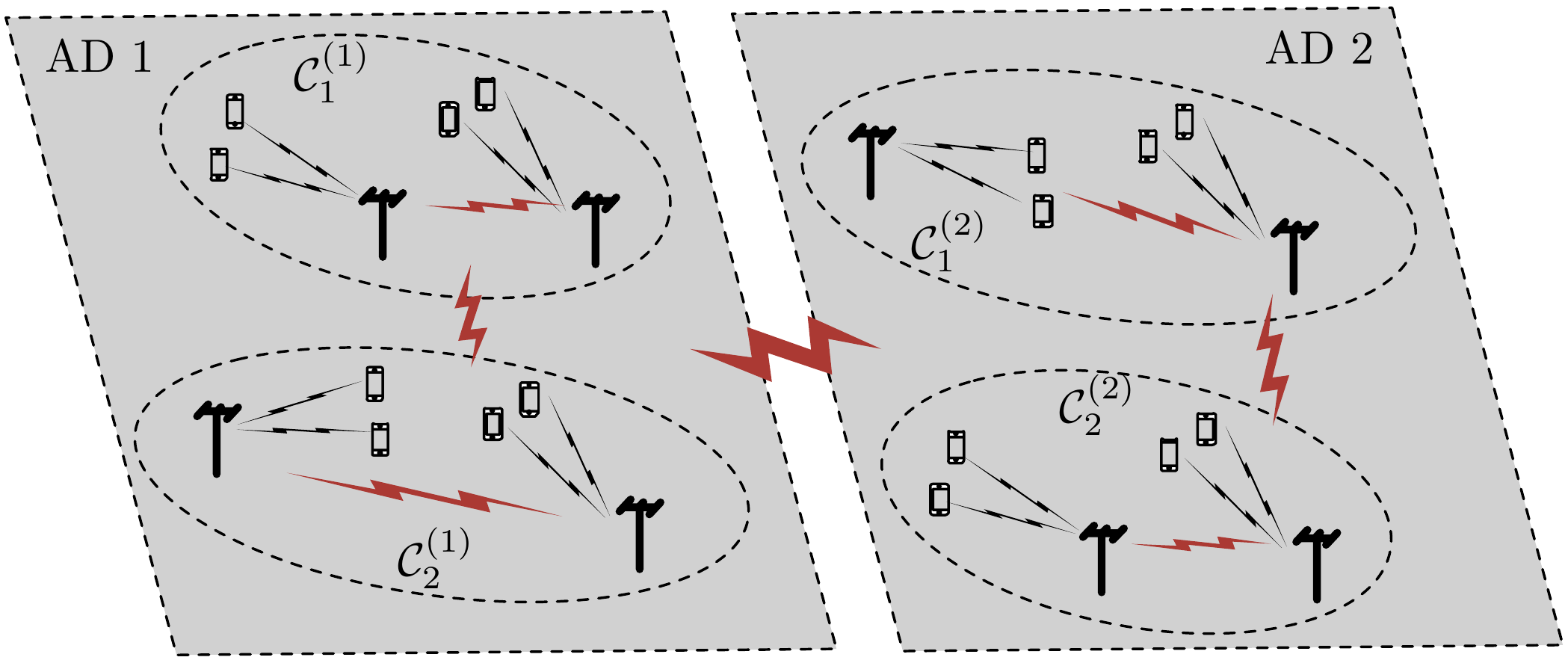} 
\caption{System-level Operation}%
\label{fig:sysmod}
\end{figure*}

We use Algorithm 1 to perform the assignment of RRHs to ADs. Then, 
there are additional coordination mechanisms deployed within each AD: cooperation within the latter is done via clusters of cooperating RRHs, that in turn are formed based on geographical distance (using algorithms such as the K-means). Moreover, all the  radio heads in the latter cluster perform coordinated beamforming (CB) (using the well known Weighted MMSE (WMMSE)~\cite{shi_wmmse_2011}), to iteratively optimize their precoders. The steps in this coordination hierarchy are detailed in the table below. 

\begin{algorithm} [h]
\begin{algorithmic}
\State \textbf{System level operation:} 
\State - Assign users to RRHs based on highest channel energy
\State - Compute coupling coefficients matrix $\bPsi$ 
\State - Run ADF algorithm to obtain RRH-to-AD assignment 
\For{ \textbf{each AD} } 
  \State - Cluster RRH into $C$ clusters using k-means (optional)
  \State - Run WMMSE (within each cluster) to optimize precoders 
 \EndFor 
\end{algorithmic}
\end{algorithm}

The result of combining the latter two `coordination mechanisms' is a hierarchical cooperation model: at the highest level, RRHs are assigned into ADs (using Algorithm 1), then, RRHs are grouped to form coordination clusters (where each is performing coordinated beamforming). The resting network structure is illustrated in Fig.~\ref{fig:sysmod}. Next, we mathematically formalized the operation and performance of the system. 

\subsubsection{Sum-rate performance}
\begin{figure*}%
\begin{align} \label{eq:sinr}
\textrm{SINR}^{(l,m)}_{i_k} = \frac{ P_t | (\bh^{(l,m)}_{i_k,i})^\dagger  \bv^{(l,m)}_{i_k}  |^2 }{  \underset{ (p,q)\neq (i,k)  }{\sum} P_t |   (\bh^{(l,m)}_{p_q,i})^\dagger  \bv^{(l,m)}_{i_k}  |^2   +  \underset{  (r, s) \neq (l,m) }{\sum}   \underset{ (p, q) }{\sum}  P_t |  (\bh^{(r,s)}_{p_q,i})^\dagger \bv^{(r,s)}_{p_q}   |^2  + (\sigma_{i_k}^{(l,m)})^2  }
\end{align}
\end{figure*}

 We assume for simplicity that each RRH is serving $J$ users, and that the size of each cluster, $|\calC_m^{(l)} |$, is the same. Moreover, let $C$ denote the number of clusters within each AD. 
 Let $\calC_m^{(l)} $ denote the $m$th cluster  ($1 \leq m \leq C$)  in AN $l$ ($1 \leq l \leq A$): then $\calC_m^{(l)} \subset \calR$ is the set of cooperating radio heads within the $l$th AD (Fig \ref{fig:sysmod}). 

For shorthand notation, we denote by $i_k$, the $i$th user ($1 \leq i \leq J$), served by RRH $k$ ($1 \leq k \leq K$).    
Then, $ \bh^{(l,m)}_{i_k,j}$  is the (MISO) channel from RRH $j$, to user $i_k$, in $\calC_m^{(l)} $. Similarly, we define $ \bv^{(l,m)}_{i_k}$ at the transmit precoder, used to serve user $i_k$, in $\calC_{m}^{(l)} $. Letting $P_t$ denote the transmit power of all radio-heads, the SINR of user $i_k$, in $\calC_{m}^{(l)}$, is given by~\eqref{eq:sinr}. Thus, treating interference as noise at the users, the \emph{achievable sum-rate} of the system is as follows, 
\begin{align} \label{eq:sumrate} 
R_{\Sigma} = \sum_{l \in \calA } \sum_{m=1}^C \sum_{i=1}^J \sum_{k=1}^{K} \ \log_2 (1  +  \textrm{SINR}^{(l,m)}_{i_k} ) 
\end{align}

Under this setup we advocate, each user is subject to residual interference from users within its coordination cluster, interference emanating from  users in other clusters (but still within the same AD), as well as interference coming from all users present in other ADs.

\subsubsection{Practical Aspects}
The matrix of coupling coefficients $\bPsi$ should available at all the ANs, prior to the start of the algorithm: the latter can be ``populated'' sequentially, by having each AD estimate all channels (both to its served users, and to users from other ADs) via training, in an orthogonal manner. 

Due to the fact that the assignment variables are binary, one only needs a low-rate control link between all the ANs. In the case where $A=2$, due to the bilinear structure of $f$, only one iteration of Algorithm 1 is required for convergence: this is quite beneficial since it keeps the communication overhead to a minimal level. Moreover, one can initialize the algorithm with several feasible solutions, and pick the best optimal solution among them.

\section{Numerical Results}
\subsection{Simulation Setup}
Both radio heads and users  are dropped uniformly within the area of interest (their positions are kept fixed throughout the simulation), where no mobility is considered.  
Then, for each simulation run, channels are generated randomly: all channels are complex i.i.d,  assumed to be slowly block fading. 

We first investigate a system where $N=16$ radio heads are deployed, each equipped with $M=4$ transmit antennas and serving $J=2$ single-antenna users (for a total of $K=32$ users). The RRHs are to be assigned to one of two ANs (i.e., $A=2$). In our simulations we assume that  within each AD, radio heads form one cluster (i.e. $C=1$), thereby forming \emph{global coordination (GC)} cluster within each AD. Then, WMMSE is employed to iteratively optimize the precoders within each cluster. Moreover, we assume that the loading factors are identical, whereby the RRH load is split equally among the two ADs, i.e. $\pmb{\beta}_k = \pmb{1}, \ \forall k $ and $\gamma_k = N/A , \ \forall k $ .

For the assignment of radio heads to ANs, we benchmark our proposed scheme, Algorithm 1,  against a \emph{randomized assignment} where radio heads are randomly assigned to each of the ANs. Despite the fact that the latter is not a good choice, it is however intended to be used as a \emph{lower bound on the performance} of our algorithm (the latter assignment  still takes into account the equal-loading constraint, and the fact that each RRH is assigned to one AD only). Moreover, for this particular (relatively small) case, we are able to find the globally optimal solution to the ADF problem, via  \emph{exhaustive} search of $(P1)$.

We also investigate two (extreme) special cases of ADF,
\begin{itemize}
\item[-] \emph{ADF based on instantaneous CSI} is the case detailed in~\eqref{eq:cij_inst} where matrix of coupling coefficients $\bPsi$ is based on instantaneous CSI: in this case, $\bPsi$ is updated every time the channel changes, and consequently, the ADs have to be recomputed at every channel realization 
\item[-] \emph{ADF based on statistical CSI}, where the coupling coefficients are given by, 
\begin{align} \label{eq:cij_stat}
c_{i,j} = 
\begin{cases} 
    \Vert \tilde{\bH}_{i,j} \Vert_F^2 + \Vert \tilde{\bH}_{j,i} \Vert_F^2 , \ \forall \ i \neq j \\
    \  0 \ , \forall i = j      
\end{cases}
\end{align}
where, analogously to $\bH_{i,j}$, we define  $\tilde{\bH}_{i,j} \in \mathbb{C}^{|\calU_j| \times M_i } $ as the matrix of \emph{pathloss factors} from the antennas of radio head $i \in \calR $, to the users of radio head $j \in \calR $. Unlike the instantaneous CSI case, here $\bPsi$ is computed at the beginning of the simulation, the  ADF performed, and remains static throughout the simulation run (since users are static). 
\end{itemize}

\subsection{Results}
\begin{figure}
    \center
	\includegraphics[width=9cm, height=5.5cm]{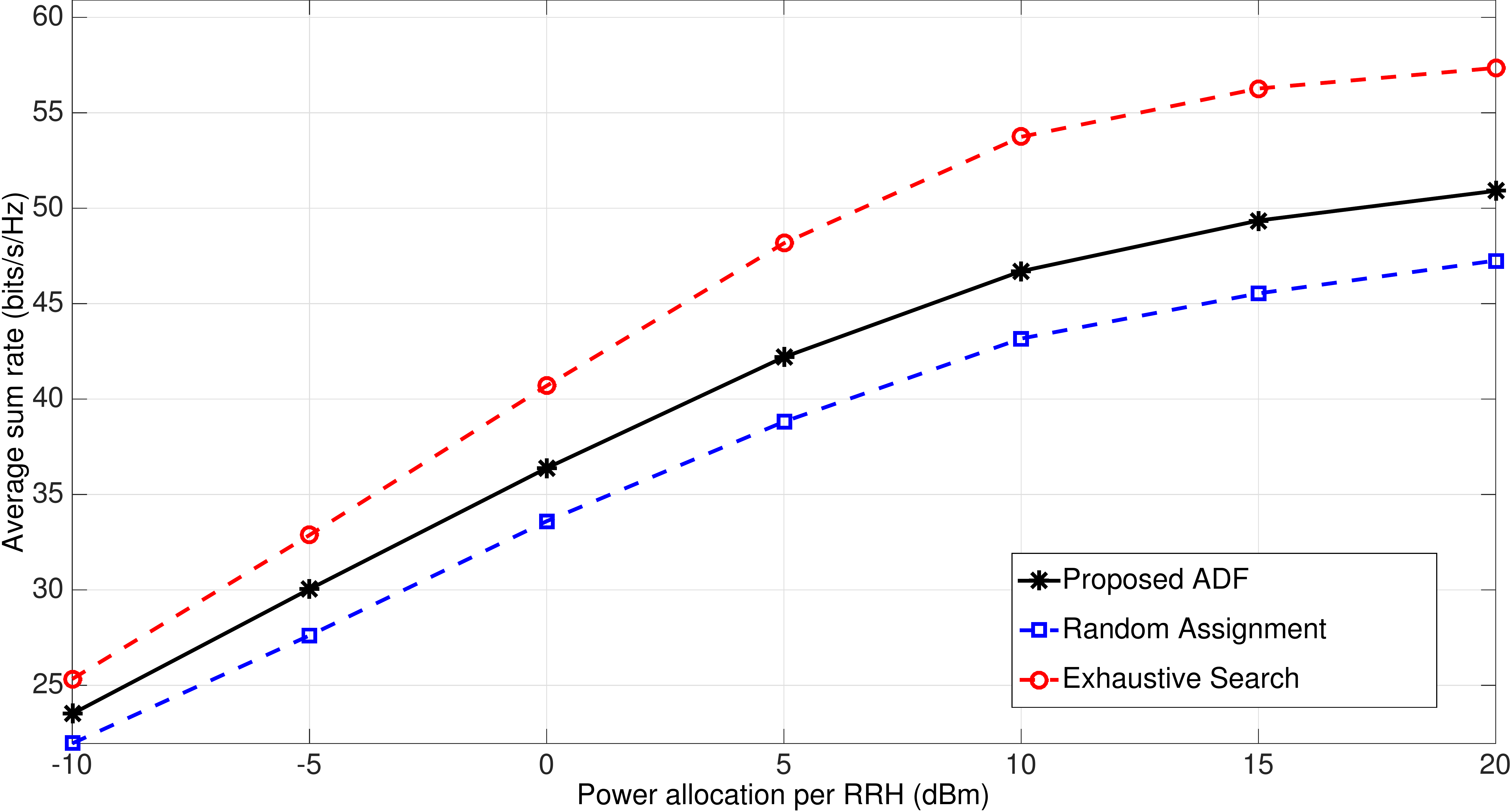}
    \vspace{.05cm}
	\caption{Average sum-rate performance using instantaneous CSI $M=4, N=16, K=32, C=1, J=2$ }
    \label{fig:sr_assign}

      \includegraphics[width=9cm, height=5.5cm]{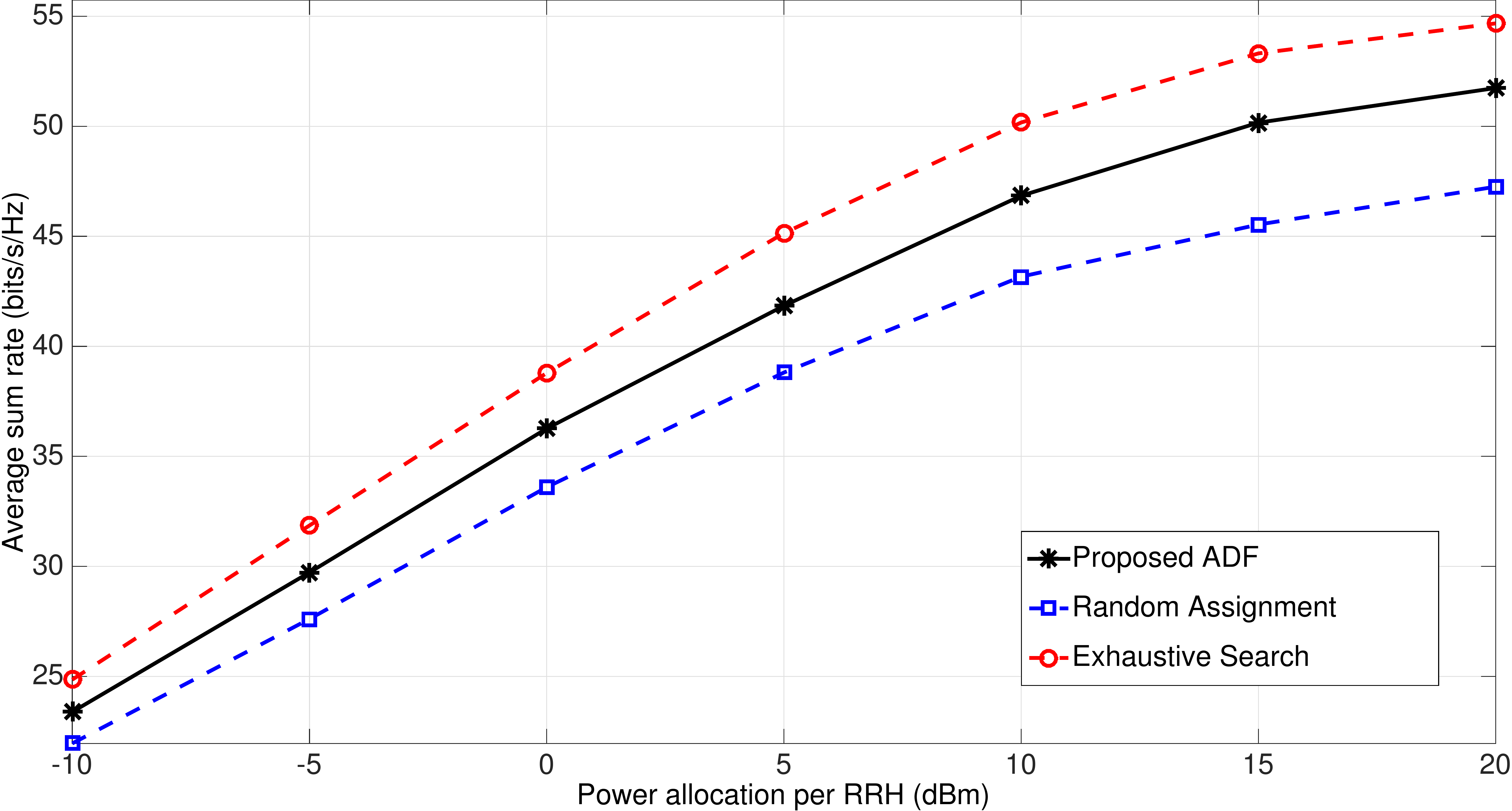}
  \vspace{.05cm}
  \caption{Average sum-rate performance using statistical CSI $M=4, N=16, K=32, C=1, J=2$ }
    \label{fig:sr_inst_vs_stat}

    \includegraphics[width=9cm, height=5cm]{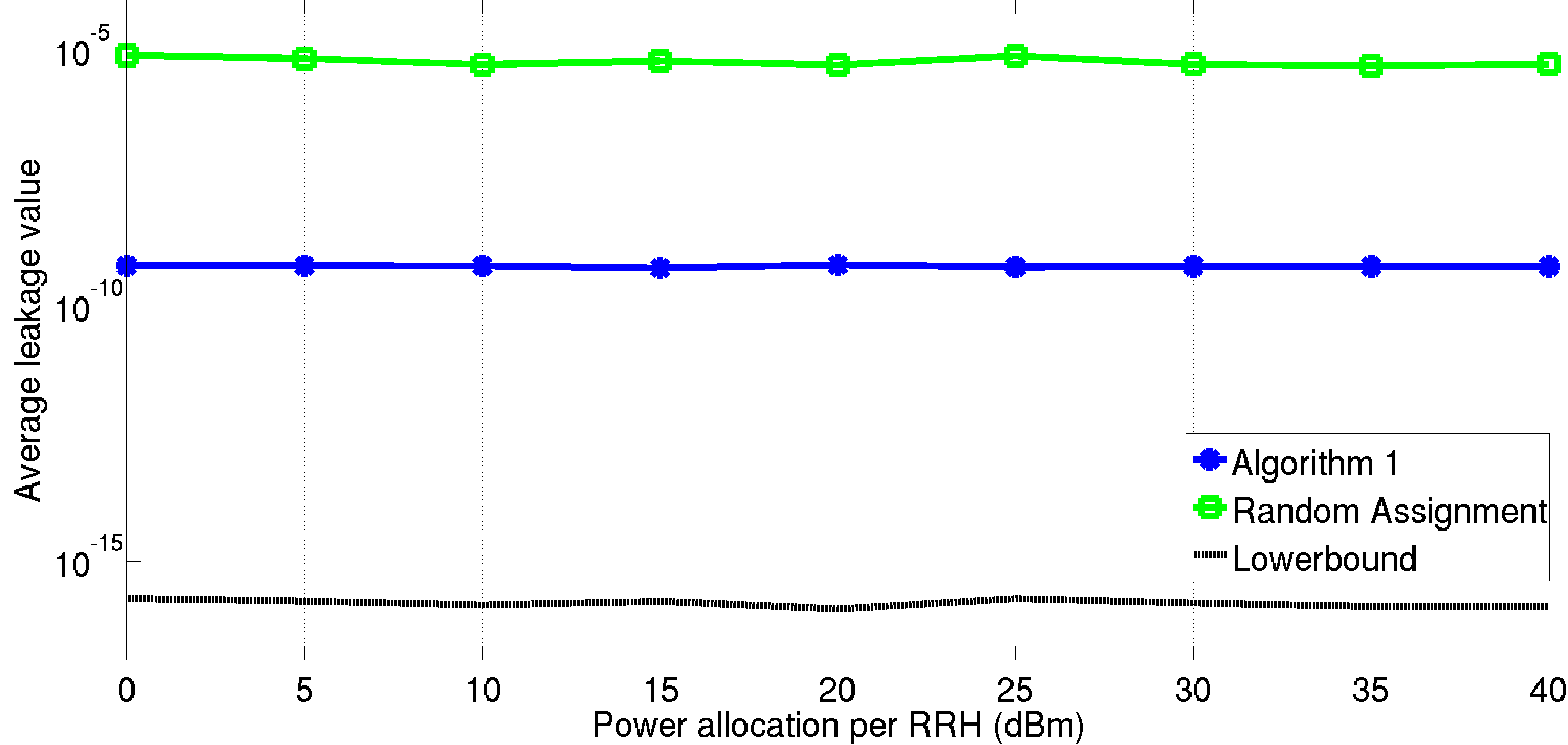}
  \vspace{.05cm}
  \caption{Average leakage value ($M=4, N=16, K=32, C=1, J=2$) }
    \label{fig:il_vs_snr}
\end{figure}

The sum-rate performance (refer to~\eqref{eq:sumrate}) that results from the above setup is shown in Fig~\ref{fig:sr_assign}, where we compare the performance of our algorithm, against both the random assignment and the optimal exhaustive search. Instantaneous CSI is assumed for all schemes. 
We can clearly see a significant performance gap between the performance of our scheme, and that of the benchmark. Moreover, looking at the performance of the exhasutive search, reveals that there are rather large gains from finding the globally optimal solution to the ADF problem. 

We next investigate the extent to which replacing Instantaneous CSI with long-term Statistical CSI, degrades the sum-rate performance of the system in question (we follow the exact same setup used for the instantaneous CSI case).  As shown in Fig~\ref{fig:sr_inst_vs_stat}, we see the same trends discussed just above. More importantly, as far as our algorithm is concerned, the degradation observed by going from instantaneous to statistical CSI, is extremely negligible. This can be exploited to massively reduce the overhead and complexity of the system (detailed in next subsection). 

 Following the same setup as above, we shed light on the behavior of the proposed ADF algorithm 1, as a function of the SNR. In Fig~\ref{fig:il_vs_snr} we show the average value of $f$ that is achieved by Algorithm 1, as well as the lower bound discussed in~\ref{sec:lb}. Though our algorithm greatly outperforms the benchmark, it is relatively far from the lower-bound (derived from the relaxed problem $(P3)$): however, we reiterate the fact that solutions such as the latter, do not correspond to feasible solutions for the ADF problem (thus are potentially not achievable within our setup).   

\subsection{Discussions} \label{sec:disc}
The numerical results all point to a good performance brought about by the application of our proposed method (both in terms of achievable sum-rates for the system, and total interference leakage between the ADs). Moreover, as shown in Algorithm~\ref{alg:1}, the communication overhead requirement among the ANs is quite negligible since it consists of binary vectors only. With that in mind, the overhead would consist of obtaining all the CSI needed to compute $\bPsi$, and propagate it to all the ANs (as often as needed). Though this might seem too high, our simulations also clearly indicated that for relatively static settings, ADF based on pathloss (i.e., location-based) has virtually the same performance as the one based on instantaneous CSI.

\section{Conclusion and Future Work}
In this work we tackled the problem of ADF in C-RAN systems by formulating it as an integer programming problem. We employ the well-known BCD framework for solving the problem and devising an iterative algorithm for that purpose. We shed light on the convergence of the algorithm, as well as its `maximal performance' via simple relaxations. Our simulations reveal that this approach promises to deliver good sum-rate performance, in typical C-RAN deployments.  

As mentioned earlier, we started to investigate more analytical aspects of the ADF problem, namely, better relaxation techniques (e.g. Lagrange Relaxation), decomposition techniques (e.g. Dantzig-Wolfe Decomposition), and dual problem analysis. The latter will provide us with insights into fundamental lower bounds for the ADF problem. 
In addition, we extended the problem formulation used here, to include both the channel and precoder effect, rather than just the channel energy (as seen in (1)). All the above issues are investigated in great detail, in our subsequent work~\cite{Ghauch_ADF_journ}. We also wish to investigate more practical scenarios, namely the so-called hybrid CSI case, consisting of a mix between instantaneous and statistical CSI.

\section{Acknowledgment }
The authors gratefully acknowledge the funding and support of their counterparts at Huawei, Finland.

\appendix
\section{}
\subsection{Proof of Lemma 1} \label{sec:app1}
Note that the following is a direct consequence of~\eqref{opt:xk_opt} 
\begin{align*}
f( \lrb{\bx_k^{(n)}} ) &\geq f(\bx_1^{(n+1)}, \bz_1^{(n)}  )  \geq f(\bx_2^{(n+1)}, \bz_2^{(n)}  ) ... \\
&\geq f(\bx_A^{(n+1)}, \bz_A^{(n)}  ) \triangleq f( \lrb{\bx_k^{(n+1)}} )   
\end{align*}
where the last equality follows from the fact that $f(\bx_A^{(n+1)}, \bz_A^{(n)}  )$ corresponds to the case where all variables $(\bx_1, ...., \bx_A)$, are updated. It follows that the sequence $ \lrb{ f(\bx_1^{(n)}, ..., \bx_A^{(n)} ) }_n $ converges monotonically to a limit point $f_0$

\addcontentsline{toc}{chapter}{Bibliography}
\bibliographystyle{ieeetr}
\bibliography{ref_hadi.bib}

\end{document}